\documentstyle[epsf]{mn}
\title[Ejecta fragments in compact supernova remnants]
{
On the evolution of ejecta fragments in compact supernova remnants
}
\author[R. Cid-Fernandes \etal]
    {R. Cid-Fernandes,$^{1,2\star}$
     T. Plewa,$^{3\star}$
     M. R\'o\.zyczka,$^{4,5\star}$
     J. Franco,$^{6\star}$
     R. Terlevich,$^{2\star}$ \cr
     G. Tenorio-Tagle$^{7\star}$
     and
     W. Miller$^{6}$\thanks{E-mail: cid@if.ufrgs.br (RCF); \newline
tomek@MPA-Garching.MPG.DE (TP); mnr@camk.edu.pl (MR); \newline
pepe@astroscu.unam.mx (JF); \newline rjt@ast.cam.ac.uk (RT);
gtt@iac.es (GTT); \newline warren@astroscu.unam.mx (WM).}
\\
    $^1$Departamento de F\'{\i}sica, CFM-UFSC, Campus universit\'ario,
Trindade, Caixa Postal 476, 88040-900 Florian\'opolis, SC, Brazil\\
    $^2$Royal Greenwich Observatory, Madingley Road, CB3~0EZ,
Cambridge, UK\\
    $^3$Max-Planck-Institut f\"ur Astrophysik,
Karl-Schwarzschild-Strasse 1, 85740 Garching bei M\"unchen, Germany \\
    $^4$Warsaw University Observatory, Al.\ Ujazdowskie 4, 00478
Warszawa, Poland \\
    $^5$N. Copernicus Astronomical Centre, ul.\ Bartycka 18, 00716
Warszawa, Poland \\
    $^6$Instituto de Astronom\'\i{}a UNAM, Apartado Postal 70-264,
04510 M\'exico D. F., M\'exico \\
    $^7$Instituto de Astrof\'\i{}sica de Canarias, 38200 La Laguna,
Tenerife, Spain\\
    }
\def\ni{\noindent}                                       
\def\etal{et al.\ }                                         

\def\gapprox{$_>\atop{^\sim}$}     

\def\kms{km\thinspace s$^{-1}$}                        
\def\ergsec{erg\thinspace s$^{-1}$}                   
\def\msun{M$_{\odot}$}                                  
\def\lsun{L$_{\odot}$}                            
\def\cms{cm\thinspace s$^{-1}$}                         
\def\percc{cm$^{-3}$}                         
\def\percucm{cm$^{-3}$}                         
\def\peryr{yr$^{-1}$}                                     

\def\ha{$H_\alpha$}                                        

\def\ltsimeq{\,\raise 0.3 ex\hbox{$ < $}\kern -0.8 em
 \lower 0.7 ex\hbox{$\sim$}\,}
\def\gtsimeq{\,\raise 0.3 ex\hbox{$ > $}\kern -0.8 em
 \lower 0.7 ex\hbox{$\sim$}\,}

\newcommand{\ET}[1]{\times 10^{#1}}             

\def\beq{\begin{equation}}                          
\def\eeq{\end{equation}}                              
\def\beqa{\begin{eqnarray}}                         
\def\eeqa{\end{eqnarray}}                             
\def\beqan{\begin{eqnarray*}}                      
\def\eeqan{\end{eqnarray*}}                          

\newcommand{\ov}[1]{\overline{#1}}                        
\newcommand{\zb}[1]{\left[ {#1} \right]}           
\newcommand{\zp}[1]{\left( {#1} \right)}               
\newcommand{\mrm}[1] {{\rm #1}}
\date
{
Accepted \hspace{5em};
Received \hspace{5em};
in original form
}
\volume{000}
\pagerange{\pageref{firstpage}--\pageref{lastpage}}
\pubyear{}
\begin{document}
\maketitle
\label{firstpage}
\begin{abstract}
We examine the evolution of inhomogeneities (fragments) of supernova
ejecta in compact supernova remnants by means of hydrodynamical
modeling and simplified analytical calculations. Under the influence
of intense post-shock cooling the fragments become strongly compressed
as they traverse the hot shocked region between the reverse and outer
shocks of the remnant. We find that the most likely outcome of the
interaction of fragments with the reverse shock and the hot shocked
region is their disruption resulting in generation of secondary
fragments. Secondary fragments arriving at the thin and dense outer
shell of the remnant give rise to brief X-ray flashes. Under suitable
conditions the primary fragments may traverse the hot shocked region
without being completely destroyed, to eventually reach the outer
shell as dense, elongated structures. Collisions of such fragments
with the shell are likely to give rise to powerful X-ray flares.
\end{abstract}
\section{Introduction}
There is growing observational evidence that the matter expelled in
supernova explosions is not uniformly distributed, but instead it
consists of clumps or fragments with a large variety of sizes, shapes
and densities. For instance, the presence of fragments has been
detected, or at least suggested, in SN 1987A (Arnett, Fryxell \&
M\"uller 1989, Lucy \etal 1989, Hanuschik \etal 1993), SN~1993J
(Spyromilio 1994, Wang \& Hu 1994), and recently in Vela (Aschenbach,
Egger \& Tr\"umper 1995, Strom \etal 1995). Fast moving metal-rich
knots with velocities exceeding $5000$~\kms\ have been identified in
Cas A (Braun, Gull \& Perley 1987, Anderson \etal 1994) and evidence
for fragmented ejecta is also found in Tycho (Seward, Gorenstein \&
Tucker 1983), Puppis A (Winkler \etal 1988), Kepler (Bandiera \& van
den Bergh 1991) and in a number of extragalactic remnants (Lasker \&
Golinowski 1991, Fesen \& Matonick 1993, Chugai 1993). This
preponderance of clumpiness suggests that ejecta fragmentation is a
common process, and its origin has been extensively studied with the
help of multi-dimensional numerical simulations (M\"uller, Fryxell \&
Arnett 1991, Fryxell 1994 and references therein). It has been found
that in a Type II supernova there are two layers, at the H/He and
He/C+O interfaces, which within minutes from core collapse develop
strong hydrodynamical instabilities resulting in the formation of
relatively dense fragments reaching, and probably exceeding, a factor
of $10$\ in density contrast with the homogeneous interfragment part
of the ejecta.

The effects caused by fragments in a SN ejecta were first explored by
McKee (1983) and Hamilton (1985) by means of analytical and simplified
numerical techniques assuming spherical symmetry (see the review by
Franco \etal 1991). They found that the presence of fragments can
increase the thermalization radius of the ejecta, and that high column
density fragments can move ahead of the main shock wave, driving a
precursor which heats the ambient medium. 2-D numerical models of
clumpy SNRs evolving in pre-existing wind cavities were studied by
Tenorio-Tagle \etal (1991). The collisions between fragments and a
pre-existing wind-driven shell can puncture the shell and induce a
rapid mixing of different gas components. In addition, the propagation
of fragments across the remnant can be partially responsible for the
observed filamentary structures and X-ray halos in these regions. More
recently, Franco \etal (1993a) explored, with an analytical approach
and 2-D numerical simulations, the effects of clumpy ejecta in the
evolution of multi-supernova remnants. As in the case of smaller
bubbles, the fragments are likely to excite and puncture the remnant
shell over time-scales of the order of $10^6$~yr, causing both
temporal and spatial variations in the X-ray emission and introducing
distortions in the shell structure.

On the other hand, two new kinds of peculiar type II supernovae (SNe)
have been observed recently, the very luminous type II radio
supernovae and the so-called `Seyfert 1 impostors' (see review by
Terlevich 1994). Both types seem to be associated with regions of
active star formation. Seyfert 1 impostors are very bright in the
optical continuum with very strong, broad \ha\ emission with no
P-Cygni profile. Both the high \ha\ luminosity and the strong radio
emission are interpreted as resulting from the interaction between the
expanding SN ejecta and a dense circumstellar medium (Chevalier 1982,
Terlevich \etal 1992, 1995, Chugai \& Danziger 1994).

Models of the interaction of SN ejecta with a high-density homogeneous
circumstellar medium (CSM) match the observed features (spectrum,
optical light curve, X~ray luminosity and emission line widths) of
these peculiar SNRs. Remnants evolving in a dense CSM ($n >
10^5$~\percucm ) reach their maximum luminosity ($L > 10^7$~\lsun) at
small radii ($R < 0.1$~pc) soon after the SN explosion ($t < 20$~yr)
while still expanding at velocities of more than $1000$~\kms (Shull
1980; Wheeler, Mazurek \& Sivaramakrishnan 1980; Draine \& Woods 1991;
Terlevich \etal 1992). Because of their small dimensions, these
remnants are designated `compact' supernova remnants (cSNRs). A key
feature of cSNRs is that radiative cooling becomes important well
before the thermalization of the ejecta is complete, resulting in the
remnant bypassing the quasi-adiabatic Sedov track. Consequently, the
shocked matter undergoes a rapid condensation behind both the leading
and the reverse shocks. Two concentric, high-density, fast-moving thin
shells are then formed.  These shells, along with the freely expanding
ejecta and a section of the still dynamically unperturbed interstellar
gas, are irradiated and ionized by the photon field produced by the
radiative shocks. The resulting photoionized gas produces broad
emission lines with line ratios that mimic those observed in the broad
line region (BLR) of Seyfert galaxies (Terlevich \etal 1992; Franco
\etal 1993b,c, Plewa 1995).

Recent observational work by Collura \etal (1994) indicates that the
cSNR in the core of the nearby starburst galaxy M~82 may be varying at
X-ray energies in less than a day. Collura \etal actually suggested
that the variable source detected with ROSAT is associated with a
massive binary, but examining the radio-maps of M~82 we found that its
position coincides with that of the dominant radio supernova in M~82
(Kronberg, Biermann \& Schwab 1985). Furthermore, Boller, Fink \&
Schaeidt (1994) have discovered a luminous Starburst galaxy with an
X-ray luminosity of $\sim 10^{42}$~\ergsec\ that has variability on
time-scales of less than a day and varies by more than a factor of two
in amplitude.  Inhomogeneities or fragments in the ejecta, we believe,
are an important ingredient in modeling rapid variability in cSNRs and
starbursts.  In this paper we focus on the evolution of clumpy
(fragmented) SN ejecta evolving in a dense circumstellar medium
characterized by the development of dynamical instabilities in the
thin, cool and dense gaseous sheets formed as a result of intense
cooling behind the shocks. In particular, we explore the possibility
that the presence of fragments in the ejecta and/or in the ambient
medium could lead to a rapid X-ray variability in cSNRs.

Section 2 presents a set of 2-D numerical simulations following the
fragments as they interact with the reverse shock and travel through
the region of hot shocked gas between reverse and forward shock
fronts. Section 3 estimates the expected X-ray variability by means of
analytical expressions and Monte Carlo simulations. Finally, section 4
summarizes our conclusions.
\section{Numerical Simulations}
\subsection{Methods, assumptions and input physics}

The hydrodynamic equations are integrated with the help of the {\sc
AMRA} code which combines the PPM method of Colella and Woodward
(1984) with the AMR (Adaptive Mesh Refinement) approach first
introduced by Berger \& Oliger (1984) and subsequently developed by
Berger \& Colella (1989). The details of the method together with the
results of test calculations are described elsewhere (Plewa \&
M\"uller 1996).  Briefly, the {\sc AMRA} code operates on a series of
nested meshes simultaneously, automatically creating higher-resolution
meshes and moving them over the base grid to follow the small-scale
features of the model. A two-dimensional version of the code is used,
able to work in spherical, cylindrical or Cartesian coordinates. For
the present simulation 
spherical coordinates ($r,\theta,\phi$) are
chosen, implying a rotational symmetry with respect to the $\theta=0$
axis.  The basic grid extends over radii $0.0\le r\le
1.3\times10^{17}$ cm and polar angles
$0^{\circ}\le\theta\le10^{\circ}$, containing 64 and 20 uniformly
distributed points in $r$ and $\theta$, respectively.

The code is allowed to generate up to three refined grids (1--3), each
one consisting of several meshes. In the $\theta$-coordinate each
refined grid provides a twofold resolution improvement with respect to
the next coarser one, while in the $r$-coordinate the resolution
improvement factors are 2, 3 and 4 for grids 1, 2 and 3,
respectively. Thus, the resolution of the finest grid is
$8.5\times10^{13}$ cm in $r$ and $6.25\times10^{-2}$ deg in $\theta$,
equivalent to that of a uniform grid of 1536$\times$160 points
extending over the whole computational domain. A new mesh is created
whenever the local relative density contrast between neighbouring
cells exceeds 0.1 or the local relative pressure contrast exceeds 0.01
(the refinement procedure is not applied to the region occupied by the
freely expanding ejecta). With that prescription for mesh generation a
speedup factor of $\sim3.5$ is achieved, meaning that a simulation
carried out on a single grid of 1536$\times$160 points would require
350 per cent of the CPU time actually used.

A reflecting boundary condition is applied at $r=0$ cm, $\theta=0$
deg, and $\theta=10$ deg, while a free outflow from the grid is
allowed for at the $r=1.3\times10^{17}$ boundary. The equation of
state is that of an ideal monoatomic gas with solar composition. 
Complete ionization is assumed, resulting in a mean molecular weight
$\mu =0.615$. Radiative, optically thin cooling of the shocked gas is
allowed for, with the cooling function taken from Plewa (1995). For
the total energy of the supernova explosion we use the standard value
of 10$^{51}$ erg, all of it in the form of kinetic energy of the
ejecta.  The explosion occurs in a uniform circumstellar medium (CSM)
with a density $n_{\mrm {CSM}}=10^7$ cm$^{-3}$ and a temperature
$T_{\mrm{CSM}}=10^4$ K.

Based on the results of Arnett (1988), we assume that the radial
density distribution of the ejecta is well represented by a $\rho (r)
\propto r^{-3}$ power law (the outer layer of the ejecta with a very
steep density gradient is neglected, as it contains only a small
fraction of the total mass of the exploding star). We further assume
that the velocity of the ejected gas is proportional to $r$, reaching
about ten thousand \kms \ at the outer edge of the ejecta.  As it is
very likely that a typical massive supernova progenitor blows off most
of its hydrogen envelope prior to the explosion, for the total mass of
the ejecta a value of 5 \msun\ is adopted.

As indicated in Section 1, the ejecta becomes fragmented soon after
the explosion due to Rayleigh-Taylor instabilities acting at the
interfaces between layers with different chemical compositions
(Arnett, Fryxell \& M\"uller 1989, Hachisu \etal 1992). Counting from
outside, the first such interface is associated with the outer edge of
the helium core, whose mass may vary between 4 \msun\ and 8 \msun\
depending on the initial (main sequence) mass of the progenitor. Thus,
with a helium-core mass of $\sim$4 \msun, the mass of the external,
relatively smooth part of the ejecta in our model is equal to $\sim$1
\msun. The properties of the fragments are poorly known at present. In
a simplified scenario one may describe the inhomogeneous part of the
ejecta as an ensemble of cloudlets with various sizes, shapes, and
densities immersed into an {\it interfragment medium} (IFM). This IFM,
apart from the large-scale radial density gradient, may be regarded as
uniform. In such a scenario, and in the absence of internal motions,
the fragments are frozen into the general diverging flow, i.e. they
expand homologously together with the IFM until they encounter the
reverse shock of the remnant. According to M\"uller, Fryxell \& Arnett
(1991), fragments as massive as 0.01 \msun\ can be found in the
ejecta. In the present simulations a less extreme case is explored,
with the mass of the fragment ranging from $3\times10^{-4}$ \msun\ to
$3\times10^{-3}$ \msun, while the whole ensemble of fragments with a
broad distribution of masses is considered in the analytical part of
the paper (Section 3).
\subsection{Initial conditions}
At $t=0$ the outer radius of the region occupied by the freely
expanding ejecta is set equal to $3.25\times10^{16}$ cm (25 per cent
of the base grid radius). Thus, our time count begins at about 2.5 yr
after the explosion. For the first 3 yr of the evolutionary time the
remnant is evolved without fragments, i.e. with the smooth IFM
only. By then, the outer shock has propagated up to $7.8\times10^{16}$
cm, while the reverse shock is located at $5\times10^{16}$ cm. Both
shocks are still in the adiabatic evolutionary phase, and between them
a hot shocked region extends, divided into two parts by a contact
discontinuity which separates the shocked ejecta from the shocked
CSM. We start the evolution of the fragments at $t=3$ yr. At this time
the mass of the shocked ejecta amounts to $\sim2.5$ \msun, which means
that not only the smooth hydrogen envelope, but also a fraction of the
inhomogeneous ejecta (consisting of IFM and fragments) have been
overtaken by the reverse shock. Thus, the fragments are inserted by
the time when they are already interacting with the reverse shock, and
we can follow the evolution of the shocked fragments and their
possible interaction with the outer dense shell.

In our simulations only one cylindrical fragment is followed. The
fragment is generated at $t=3$ yr in the freely expanding ejecta, with
its front part almost touching the reverse shock, and its shape
approximated by a slightly elongated cylinder whose base is
perpendicular to the radial direction (i.e. to the direction of
motion). We define the $\epsilon$ as the {\it density contrast\/}
between the fragment and the interfragment media. The generation
procedure simply increases the local density of the ejecta within the
cylinder by a constant factor $\epsilon$, while the velocity and
specific internal energy of the ejecta remain unchanged. The radius
$R_{\mrm {f}}$ and 
 length $l_{\mrm{f}}$ of the cylinder are the
same for all simulations, amounting to $1.5\times10^{15}$ cm and
$4.5\times10^{15}$ cm, respectively. With these dimensions, the range
of fragment 
masses defined in the preceding section corresponds to 
an $\epsilon$-range of 3 to 30.  The velocity of the front
part of the fragment amounts to $\sim 4500$ \kms \ (roughly equal to
the largest velocity of the unshocked ejecta).
\subsection{Results}
Four different simulations have been performed, and we refer to them
as cases A--D. Case A corresponds to 
(no fragment), and
case B, C and D to $\epsilon=3$, 10, and 30, respectively. In all
cases the evolution is followed up to $t=11$ yr, long past the thin
shell formation phase.

In case A, which serves as a reference case (not shown), the remnant
evolves with a smooth ejecta without fragments. The shell begins to
form at $t\sim6$ yr at a distance of $\sim2.5\times10^{16}$ from the
outer shock front. Within less than 1 yr all the gas contained between
the original shell formation site and the outer shock cools down and
collapses into a very thin layer of shocked CSM (see Franco \etal
1994). This process occurs very rapidly and has been termed {\it
catastrophic cooling} (see Plewa 1995 and references therein). Given
the improved spatial resolution allowed by the adaptive method, {\it
this is the first time} that the initiation of catastrophic cooling,
and its subsequent migration as a cooling wave toward the outer shock,
has been observed in 2-dimensional simulations of supernova
remnants. The cooling wave also propagates towards the contact
discontinuity, condensing the shocked CSM onto the thin shell and
causing the distance between the shell and the contact discontinuity
to decrease. The hottest gas, which was the one shocked at the
earliest moments, remains hot for a long time and even at the end of
the simulation the remnant still contains an appreciable amount of
high temperature gas. Our model resolution provides an adequate
{\it qualitative} description of the process, but most values and details
cannot be completely resolved. Thus, the present results only provide
a reasonable guideline of the expected evolution.

The basic evolutionary features of this case are also followed by the
other three cases because the presence of a fragment in the flow does
not alter significantly the {\it global} evolution of the remnant. In
particular, the shell formation stage always proceeds in the same way
and with the same time-scales. The main difference among models is in
the details of the fragment evolution, and in the emission and local
structures generated at the sites of interaction.

Snapshots of cases B--D taken at $t=4$ yr are shown in Fig.
\ref{fig:3cases}. The stratified region extending up to
$5.4\times10^{16}$ cm from the center of the remnant contains the
freely expanding (unshocked) ejecta. The sharp density contrast at
$r\approx5.4\times10^{16}$ cm marks the momentary location of the
reverse shock front. The hot shocked region extends to the right of
the reverse shock, where a sharp and smooth right-hand boundary at
$r\approx8.6\times10^{16}$ cm marks the location of the outer
shock. The contact discontinuity separating the shocked ejecta from
the shocked CSM is visible as a light-grey patch situated at
$r\sim6.8\times10^{16}$ cm, roughly in the middle of the hot shocked
region.
%
%
\begin{figure*}  
\begin{minipage}{17.5cm}
\protect\centerline{
\vspace{22 cm}
                   }
\caption{
Density distributions in cases B--D (top to bottom) at $t=4$ yr. In
case B ($\epsilon=3$) the fragment is entirely shocked, while the in
case D ($\epsilon=30$) the shocks driven into the fragment have swept
about 75 per cent of its mass. The figure was obtained by reflecting
all meshes with respect to the $\theta=0$ axis, i.e. the base grid
extends over half of the area shown. Small asymmetries in plot details
visible far away from the symmetry axis are artifacts of the plotting
program.
\label{fig:3cases}
}
\end{minipage}
\end{figure*}
%
%
\begin{figure*}  
\begin{minipage}{17.5cm}
\protect\centerline{
\vspace{22.5 cm}
                   }
\caption{
The evolution of case D. Density distributions in the outer part of
the remnant are shown at $t=5$, 6 and 7 yr.  The figure was obtained
by reflecting all meshes with respect to the $\theta=0$ axis, i.e. the
base grid extends over half of the area shown. Small asymmetries in
plot details visible far away from the symmetry axis are artifacts of
the plotting program.
\label{fig:caseD}
}
\end{minipage}
\end{figure*}
The smooth indentation in the middle of the reverse shock, visible in
all three panels of Fig. \ref{fig:3cases}, is the result of a
perturbation caused by the presence of the fragment (the shock slows
down at the edges of the fragment). 
After the fragment has been
entirely engulfed by the reverse shock, a conical self-reflection of
the reverse shock is initiated, leading to a complicated pattern of
secondary shocks which can be best seen in the first panel of
Fig. \ref{fig:3cases}, and in all three panels of Fig. \ref{fig:caseD}
(see also the detailed description in Klein \etal 1994). Secondary
shocks are also driven into the rear side of the fragment. As all
shocks sweeping through the fragment are strongly radiative (see
Section 3.1), the re-expansion phase described by Klein \etal (1994)
is never reached. Instead, the fragment is compressed to the limits
imposed by the resolution of the grid creating thin shells behind the
penetrating shocks. Thus, aside from the Kelvin-Helmholtz
instabilities described by Klein \etal, the shocked gas is also 
subjected to the thin shell instability and the obvious outcome of the
interaction is the breakup of the original fragment into smaller and
denser secondary fragments.  At $t=4$ yr (Fig. \ref{fig:3cases}) the
breakup has only begun. In all three panels the fragment can still be
recognized as a single object, but the effects of ablation (see also
Kimura \& Tosa 1991) and secondary fragment stripping are already
visible.  The main body of the fragment clearly shows the expected
(see Section 3.1) dependence of the mean velocity on the initial
density contrast (compare the positions of the fragment in all panels;
compare also the strength of the bow shock in front of the fragment).
The secondary fragments are very dense and move like solid particles
inside the hot region, with the velocities of the flow at the places
and moments of their creation (in particular, secondary fragments both
overtaking the front edge of the primary fragment and lagging behind
it may be seen in the lower two panels of Fig. \ref{fig:3cases}).

The subsequent evolutionary stages of case D are shown in
Fig. \ref{fig:caseD}. In all three panels only the magnified hot
shocked region is displayed, bordered by the reverse shock (with a
clear conical reflection pattern) on the left-hand side, and by the
outer shock on the right-hand side. At $t=5$ yr the formation of
secondary fragments is completed. Being much 
denser than the
shocked IFM, the fragments move essentially balistically across the
hot shocked region. At later evolutionary stages, the effects of
ablation and further fragmentation can be seen. At the same time,
however, some of the secondary fragments seem to be merging into
larger entities (e.g. see the triple fragment at the symmetry axis and
at $r\approx9\times10^{16}$ cm in Fig. \ref{fig:caseD}).

As mentioned above, between $t=6$ yr (middle panel) and $t=7$ yr
(lower panel), a significant fraction of the shocked CSM behind the
outer shock undergoes catastrophic cooling and collapses into a very
thin and dense shell. Immediately after the dense shell has formed,
the first fragments begin to collide with it.  The results of the
collisions are shown in Fig. \ref{fig:Lgrid}, where the radiative
activity of the modeled part of the remnant is illustrated. Obviously,
given the large velocities involved (specially in the case of shocks
faster than $\sim2000$ \kms ) most of the thermalized kinetic energy
is emitted in the form of X-rays (Plewa 1995), so that the X-ray and
total light curves are very similar.

The light curve of case A (Fig. \ref{fig:Lgrid}a) is qualitatively
much the same 
as the light curves obtained by Plewa (1995) for
plane-parallel radiative shocks. For the first five years the
luminosity of the remnant increases steadily due to the increasing
cooling of the shocked CSM. At $t=5$ yr the first mass elements of the
shocked CSM reach the minimum of the cooling curve and enter the
catastrophic cooling phase, in which the cooling rate increases with
decreasing temperature. As a result, an X-ray/UV flare is generated,
followed by a deep luminosity minimum between $t=5$ yr and $t=6$
yr. The second flare at $t=7$ yr is originated when the cooling wave
arrives at the outer shock, and strong secondary shocks are driven
into the cool CSM, completing the formation of 
a thin and dense shell
(see Plewa 1995 for a detailed description of this process). The
differences in luminosities between the reference case A and the ones
with fragments are illustrated in Fig. \ref{fig:Lgrid}b. The first,
low-amplitude peak is associated with the catastrophic cooling regime
behind the shocks driven into the fragment. The total energy emitted
in this peak is mainly dependent on the fragment density, with the
largest luminosity corresponding to the largest $\epsilon$. The
bombardment of the shell with secondary fragments begins at
$t\approx6.25$ yr (corresponding to the second significant peak in
Fig. \ref{fig:Lgrid}b). As the distance between the formation site of
the secondary fragments and the shell is equal to $\sim2\times10^{16}$
cm (see Fig. \ref{fig:3cases}), one may estimate the speed of the
fastest fragments arriving at the shell at $\sim 3200$ \kms , which is
below the initial fragment velocity (some $4500$ \kms ). In contrast
with the first peak, the total energies emitted in the second and
subsequent peaks are correlated with the kinetic energy of the
fragment impacting the shell.
%
%
\begin{figure*}
\begin{minipage}{17.5cm}
    \protect\centerline{
    \epsfxsize=17.5cm\epsffile[40 390 570 650]{figures/fig3.ps}
     }
\caption{
    (a) Light curve of the remnant with homogeneous ejecta, where
$L_{\mrm{grid}}$\ is the total energy radiated from the fraction of
the remnant subtended by the grid. Luminosities for the whole remnant
should be $132$\ times larger than $L_{\mrm{grid}}$. (b) Differences
in X-ray luminosities between remnants with homogeneous and
inhomogeneous ejecta. Thin solid: $\epsilon=3$, dotted: $\epsilon=10$,
heavy solid: $\epsilon=30$.
    \label{fig:Lgrid}
    }
\end{minipage}
\end{figure*}
\subsection{Discussion}
Our simulations show that an isolated fragment interacting with the
reverse shock (and the hot shocked region) of a cSNR evolves through
several distinct phases. Because all shocks are strongly radiative,
the shocked fragment is compressed into a cool, dense, and 
unstable structure.  Subsequently, ablation and breakup of the
unstable structure results in the formation of secondary fragments
with very high densities. Later, these secondary fragments collide
with the outer shell of the remnant, giving rise to brief X-ray
flashes. This general picture is not very sensitive to the fragment
sizes, shapes, or initial density contrasts. It is conceivable, that
at later evolutionary stages (when the distance between the outer
shell and the reverse shock begins to decrease due to radiative
cooling of the shocked ejecta), large-$\epsilon$ fragments might
collide directly with the shell before breakup (see Tenorio-Tagle 1994
for preliminary results of the evolution with $\epsilon=100$). These
collisions should result in powerful X-ray flares.

Our results provide an adequate qualitative description of the
fragment evolution and interaction, but the details of the strong
cooling phase and the thickness of the outer shell are not completely
resolved. In fact, the shell may be more than two orders of magnitude
thinner than our resolution limit (see Section 3.2). Similarly, both
the parameters of secondary fragments (specially sizes and densities)
and their trajectories should be considered as rough, qualitative
approximations. In general, an increase in numerical resolution would
result in denser and thinner shocked layers. If the number of
secondary fragments were not changing with increasing resolution, the
characteristic time-scales of X-ray flashes would become shorter, and
their amplitudes higher. Unfortunately, on the basis of the present
simulations alone we are not able to decide if this assumption is
realistic, i.e. if strong cooling combined with hydrodynamical
instabilities results in formation of a few secondary fragments rather
than in a complete destruction of the primary fragment. Nonetheless,
it is expected that some fragments can collide with the shell before
their complete destruction.  The principal conclusion of our
simulations is that compact supernova remnants with fragmented ejecta
should have a {\it bombardment phase}, in which dense clumps of cool
gas collide at high velocities with the dense and cool outer shell
creating X-ray flares.
\section{X-ray variability of cSNRs with fragmented ejecta}
The shell bombardment phase can last several years during which the
remnant may be highly variable in the X-rays. Rapid X-ray variability
is a direct observable diagnostic of the presence of fragments in
cSNRs (see Cid Fernandes \& Terlevich 1994 for the effects of the
shell distortion on the emission line profiles). It is therefore
important to estimate the variability properties one might expect to
find in such objects. Since many different fragments may be involved
simultaneously in interactions giving rise to X-ray flashes, a direct
numerical approach employing high resolution simulations is beyond our
present capabilities.  In the following, a statistical approach
towards the X-ray variability of cSNRs will be applied, based on a
simplified analytical expressions for the fragment evolution and the
resulting interaction with the thin outer shell of the remnant.
\subsection{The evolution of fragments in the hot shocked region of the remnant}
Like in the preceding section, we assume that the interfragment
gas is homogeneous. The thermal pressure of the shocked IFM
can be evaluated from the conditions of mass and
momentum conservation across the reverse shock front. If the IFM
entering the shock has a density $\rho_{\mrm{ifm}}$\ and a negligible
pressure, and if it is processed by the reverse shock at a velocity
$u_{\mrm{ifm}}$, then the post-shock pressure is

\beq
\label{eq:phsr}
P_{\mrm{hsr}} = \xi_{\mrm{ifm}}^{-1} \rho_{\mrm{ifm}}^{} u_{\mrm{ifm}}^2
\eeq

\ni  where $\xi_{\mrm{ifm}} = 4/3$\ for an adiabatic shock or $1$\ for an
isothermal shock. 
A rough, order of magnitude estimate for $u_{\mrm{ifm}}$\ can be  obtained
using the planar two-streams approximation, which yields

\beq
\label{eq:frag_IFM_shock_vel}
u_{\mrm{ifm}} \approx \xi_{\mrm{ifm}} \frac{a_{\mrm{ifm}}}{1 + a_{\mrm{ifm}}} v_{\mrm{ej}},
\eeq

\ni where $v_{\mrm{ej}}$\ is the velocity of the ejecta at the
location of the reverse shock, 
$a_{\mrm{ifm}} \equiv (n_{\mrm{CSM}} \xi_{\mrm{CSM}} / 
n_{\mrm{ifm}} \xi_{\mrm{ifm}})^{1/2}$, and $n_{\mrm{ifm}}$
is the number density of the 
interfragment medium. For an adiabatic
shock with $n_{\mrm{CSM}} \approx n_{\mrm{ifm}}$, the reverse
shock processes the IFM at a speed of about $2v_{\mrm{ej}}/3$, and
the shocked gas streams away from the front at a speed of
$u_{\mrm{ifm}}/4$.

$P_{\mrm{hsr}}$ defined by equation (\ref{eq:phsr}) is the pressure that
the fragments meet after passing through the reverse shock. From the
jump conditions applied to a fragment that traverses the hot shocked
region one easily obtains the strength of the {\it lateral} shocks
driven into the lateral parts of the fragment:

\[
\xi_{\mrm{f}}^{-1} \rho_{\mrm{f}} u_{f, l}^2 =
  P_{\mrm{hsr}} =
  \xi_{\mrm{ifm}}^{-1} \rho_{\mrm{ifm}} u_{\mrm{ifm}}^2
\]

\ni For a fragment $\epsilon$\ times denser than the IFM, the
relationship between $u_{\mrm{f,l}}$\ and $u_{\mrm{ifm}}$\ becomes

\beq
\label{eq:vel_rs_frag}
u_{\mrm{f,l}} =
  \zp{ \frac{\xi_{\mrm{f}}}{\xi_{\mrm{ifm}}} }^{1/2}
  \frac{u_{\mrm{ifm}}}{\epsilon^{1/2}}
\eeq

\ni As $u_{\mrm{f,l}} < u_{\mrm{ifm}}$, the shocks processing
the lateral walls of the fragment are weaker than the reverse shock
propagating through the IFM. The correspondingly lower post-shock
temperature and higher post-shock density inside the fragments imply
that the fragments cool faster than the IFM. In particular,
high-$\epsilon$ fragments may cool so rapidly that their internal
shocks are practically isothermal ($\xi_{\mrm{f}}=1$). At the same
time, since $u_{\mrm{ifm}}$ is typically of the order of several
thousand \kms , the reverse shock remains adiabatic
($\xi_{\mrm{ifm}}=4/3$).

To illustrate matters, let us consider the case of a reverse shock
processing the IFM at a speed of $7000$~\kms, and assume that the
density in the ejecta just ahead of the shock is
$10^7$~\percc. Neglecting expansion effects, the cooling time for gas
with solar abundances which enters a strong shock at a velocity
$v_{\mrm{s}}$\ is (Franco \etal 1993b):

\beq
\label{eq:cooling_time}
\tau_{\mrm{cool}} \approx
    \left\{
      \begin{array}{ll}
        0.2 \zp{ v_8 / n_7 }~\mbox{ yr}   & \mbox{ if}~v_8 > 1.6 \\
        0.1 \zp{ v^3_8 / n_7 }~\mbox{ yr} & \mbox{ if}~v_8 < 1.6
      \end{array}
    \right.
\eeq

\ni where $v_8 \equiv v_{\mrm{s}}/ 10^3$~\kms\ and $n_7$\ is the
pre-shock density in units of $10^7$~\percc, and the two regimes for
$v_{\mrm{s}}$\ larger or smaller than $1600$~\kms\ are defined by
the `turning point' in the cooling function, at $T \approx 3\ET{7}$~K.
Hence, while the IFM takes about $1.4$~yr to cool, an $\epsilon =
100$\ fragment is processed by shocks of only $\sim 600$~\kms\ (using
$\xi_{\mrm{f}} = 1$\ and $\xi_{\mrm{ifm}} = 4/3$), and it
radiates its post-shock thermal energy in just a few hours, while an
$\epsilon = 30$\ fragment would have $u_{\mrm{f,l}} \sim
10^3$~\kms\ and a cooling time of a few days.

Similar considerations apply to the shock which processes the {\it
front} part of the fragment.  Because of its forward motion through
the hot shocked region, the front part of the fragment is exposed to
an extra pressure $P_{\mrm{rel}} = \rho_{\mrm{hsr}}
v_{\mrm{rel}}^2$, where $v_{\mrm{rel}}$\ is the relative velocity of
the shocked fragment with respect to the shocked IFM, and 
$\rho_{\mrm{hsr}}$\ can be approximated by the IFM's post shock
density ($4 \rho_{\mrm{ifm}}$\ for $\xi_{\mrm{ifm}} = 4/3$). The
reverse shock therefore processes the front part of fragment at a
speed

\beq
\label{eq:vel_rs_frag_FRONT}
u_{f,f} = u_{\mrm{ifm}} 
          \zb{ \frac{1}{\epsilon} \frac{\xi_{\mrm{f}}}{\xi_{\mrm{ifm}}} 
                \zp{ 1 + \frac{P_{\mrm{rel}}}{P_{\mrm{hsr}}} } }^{1/2}
\eeq

The relative velocity $v_{\mrm{rel}}$\ can be written as

\[
v_{\mrm{rel}} = \frac{u_{\mrm{ifm}}}{\xi_{\mrm{ifm}}} - \frac{u_{f,f}}{\xi_{\mrm{f}}}
\]

\ni Replacing this relation into (\ref{eq:vel_rs_frag_FRONT}) we find
that $u_{f,f}$\ is between $20$\ and $70$\ per cent larger than
$u_{\mrm{f,l}}$\ for $\epsilon$\ between $10$ and $100$. The
resulting cooling times would therefore be similar to those derived
for the lateral shocks (i.e., of the order of days or less for
$\epsilon$\gapprox$30$), still much shorter than the time the fragment
needs to traverse the hot shocked region ($\sim2$ yr in our numerical
models, see Section 2.3). Let us note that both the lateral and the
front shocks are acting simultaneously, which means that the front
shock propagates along the lateral walls through the gas that has been
already shocked by lateral shocks, i.e. its velocity at the lateral
walls of the fragment must be lower than the above estimate.
Therefore, the above estimate for $u_{f,f}$ should be regarded as an
upper limit for the effective (mass-weighted) front-shock velocity.

Finally, $v_{\mrm{ifm}}^\star$ and
$v_{\mrm{f}}^\star$, the velocities of the shocked IFM and the
shocked fragment {\it relative to the centre of the remnant} are:

\beqan
\label{eq:RELATIVE_velocities}
\begin{array}{lll}
v_{\mrm{ifm}}^\star
  & = &
   v_{\mrm{ej}} - \frac{u_{\mrm{ifm}}}{\xi_{\mrm{ifm}}} =
    v_{\mrm{ej}} \frac{1}{1 + a_{\mrm{ifm}}} \\
v_{\mrm{f}}^\star
  & = &
    v_{\mrm{ej}} - \frac{u_{\mrm{f}}}{\xi_{\mrm{f}}} =
   v_{\mrm{ej}} \zb{ 1 - \zp{ \frac{\xi_{\mrm{ifm}}}{\epsilon\xi_{\mrm{f}}} }^{1/2}
    \frac{a_{\mrm{ifm}}}{1 + a_{\mrm{ifm}}} }, 
\end{array}
\eeqan

\ni where, for simplicity, we have assumed that
$u_{f,f}=u_{\mrm{f,l}}\equiv u_{\mrm{f}}$.  For
$a_{\mrm{ifm}} = 1$, $\xi_{\mrm{ifm}} = 4/3$, and
$\xi_{\mrm{f}} = 1$, these expressions yield
$v_{\mrm{ifm}}^\star = v_{\mrm{ej}}/2$\ and
$v_{\mrm{f}}^\star = v_{\mrm{ej}}[1-(3 \epsilon)^{-1/2}]$. The
fraction of the fragment kinetic energy which is thermalized and
radiated away upon the interaction of the fragment with the hot
shocked medium is $1 - (v_{\mrm{f}}^\star/v_{\mrm{ej}})^2$. This
loss amounts to only $10$\ to $30$\ per cent for $\epsilon$\ between $100$\
and $10$, so that an appreciable energy is still available for the
fragment-shell collision.

Although the difference in lateral and front shock velocities is not
too large (see the preceding paragraphs) the fragment is squeezed
predominantly in the direction of motion, becoming more and more
flattened. At the same time it is 
subjected to several instabilities
which, as we know from the numerical simulations, result in the generation
of secondary fragments. Since low-$\epsilon$ fragments may be
completely destroyed and mixed with the shocked IFM, in the following
we shall focus on fragments with $\epsilon>10$, whose secondary
fragments are very likely to collide with the outer shell at
velocities of several thousand \kms (high-$\epsilon$ primary fragments
may even reach the outer shell before the disruption process is
completed; see Section 2.4).

An isothermal shock propagating through the fragment with a Mach
number ${\cal M}_{\mrm{f}}$ compresses the gas by a factor of ${\cal
M}_{\mrm{f}}^2$.  Assuming that the sound speed in the unshocked
fragment is $10$~\kms, for the example with $u_{\mrm{ifm}} =
7000$~\kms\ discussed above one gets $u_{\mrm{f}} \approx 600$~\kms\
for $\epsilon = 100$, or $10^3$~\kms\ for $\epsilon = 30$, with
corresponding compression factors between $3600$\ and $10000$.  Thus,
fragments with a pre-shock density of
$\approx 10^9$~\percc\ and a length (in the direction of motion) of
$\approx 10^{14}$~cm may reach densities $\sim
10^{12}$--$10^{13}$~\percc, while their lengths decrease to $\sim
10^{10}$--$10^{11}$~cm (Fig. \ref{fig:Scheme}a).
\subsection{Fragment-shell collisions}
The cooled and condensed (primary or secondary) fragments eventually
reach the outer shell of the remnant and collide with it at high
velocities to produce hard X-ray flares. A collision between a
fragment and the shell generates two new shock waves: a forward shock
moving into the shell and a reverse shock moving into the
fragment. 
The properties of these shocks, and their associated radiative bursts,
can be estimated using the two-streams approximation. Again, assuming
planar shocks, the velocities of the forward and reverse shocks are
given by

\beqan
v_{\mrm{fs}} & = & \xi_{\mrm{fs}} \frac{1}{(1 + a)} v_{\mrm{f}} \\
v_{\mrm{rs}} & = & \xi_{\mrm{rs}} \frac{a}{(1 + a)} v_{\mrm{f}} 
\eeqan

\ni where $v_{\mrm{f}}$\ is the velocity at which the fragment impacts
the shell, 
$a \equiv (n_{\mrm{s}} \xi_{\mrm{fs}} / n_{\mrm{f}}
\xi_{\mrm{rs}})^{1/2}$\ and $n_{\mrm{f}}$\ and $n_{\mrm{s}}$\ are
fragment and shell density, respectively. The shock crossing times for
the fragment and the shell are respectively

\beqan
\tau^{\mrm{s}}_{\mrm{xing}}
  & = &
    \frac{l_{\mrm{s}}}{v_{\mrm{fs}}} = \frac{1}{\xi_{\mrm{fs}}} (1 + a) \frac{l_{\mrm{s}}}{v_{\mrm{f}}}
\eeqan

\ni and

\beqan
\tau^{\mrm{f}}_{\mrm{xing}}
  & = &
    \frac{l_{\mrm{f}}}{v_{\mrm{rs}}} = 
    \frac{1}{\xi_{\mrm{fs}}} \frac{(1 + a)}{a} \frac{l_{\mrm{f}}}{v_{\mrm{f}}} \\
\eeqan

\ni where $l_{\mrm{f}}$\ and $l_{\mrm{s}}$\ represent the thickness of
the fragment and the shell, respectively. Both forward and reverse
shock waves remain radiative until one of them exits the high-density
regions, i.e., until either the reverse shock overruns the fragment or
the forward shock reaches the shell outer boundary.  To compute the
flare time-scale we shall approximate the duration of the collision
(i.e., the time during which both shocks remain active) by the minimum
of the two crossing times: $\tau_{\mrm{coll}} =
\min(\tau^{\mrm{f}}_{\mrm{xing}},\tau^{\mrm{s}}_{\mrm{xing}})$.  Note
that $\tau_{\mrm{coll}}$\ is {\it not} the duration of the flare. The
flare time-scales can be estimated assuming that the radiative burst
starts immediately after the shock passage and that the shocked gas
remains radiative for a cooling time. In this case we have
$\tau^{\mrm{f}}_{\mrm{flare}} \approx \tau_{\mrm{coll}} +
\tau^{\mrm{f}}_{\mrm{cool}}$\ for the fragment, and similarly for the
shell. Seen by an external observer, flares may be further stretched
in time if the light travel time $\tau_{\mrm{light}} = 2 R_{\mrm{f}}
\sin{\theta} / c$\ between the near and far sides of the flaring
regions is comparable to the crossing and cooling times. Here
$\theta$\ is the angle at which the collision occurs, measured with
respect to the observer's line of sight to the centre of the remnant,
and $R_{\mrm{f}}$\ is the characteristic size of the fragment in the
direction parallel to the shell. For simplicity, we assume that the
fragment approaching the shell has the shape of a flat cylinder, and
we shall regard $R_{\mrm{f}}$\ as the radius of that cylinder. In
general, the observed duration of the flare is

\beq
\label{eq:tau_flare}
\tau^{(f,s)}_{\mrm{flare}} \approx
    \tau_{\mrm{coll}} + \tau^{(f,s)}_{\mrm{cool}} + \tau_{\mrm{light}}
\eeq

\ni and the duration of the total flare is the maximum among
$\tau^{\mrm{f}}_{\mrm{flare}}$\ and
$\tau^{\mrm{s}}_{\mrm{flare}}$. The thermal energy generated during
the fragment-shell collision is quickly radiated away, and the total
energy contained in fragment and shell flares is

\beq
\label{eq:E_flare}
E_{\mrm{flare}} =
    E^{\mrm{f}}_{\mrm{flare}} + E^{\mrm{s}}_{\mrm{flare}} =
    E^{\mrm{f}}_{\mrm{kin}} \frac{a}{1 + a} \min(\eta,1)
\eeq

\ni where $\eta \equiv \tau^{\mrm{s}}_{\mrm{xing}} /
\tau^{\mrm{f}}_{\mrm{xing}} = a\,l_{\mrm{s}}/l_{\mrm{f}}$.  The
maximum flare energy corresponds to the case $a\rightarrow \infty$, in
which the fragment is completely halted at the shell and its kinetic
energy is all radiated away. The spectral range in which the flare
radiates is determined by the post-shock temperature, which for a
shock velocity $v_{\mrm{s}}$\ can be calculated from

\beq
\label{eq:Shock_Temperature}
T_{\mrm{s}} \simeq 1.2 \zp{ \frac{ v_{\mrm{s}} }
                                 { 10^8 \mrm{cm}\mrm{s}^{-1} }
                          }^2 \mbox{keV}
\eeq

%
%
\begin{figure}
    \protect\centerline{
    \epsfxsize=8.7cm\epsffile[48 240 561 550]{figures/fig4.ps}
     }
\caption{Schematic representation of the evolution of a dense fragment
inside a cSNR. (a) Interaction with the reverse shock and the hot
shocked region, leading to the formation of a thin and dense pancake
(for clarity, the lateral shocks are not shown). (b) Collision with
the outer shell. Dark areas indicate shocked regions. 
\label{fig:Scheme}}
\end{figure}

To clarify the analytical model, the evolution of a fragment as it
interacts with the reverse shock and later collides with the shell is
schematically illustrated in Fig.~\ref{fig:Scheme}. In this particular
example, the fragment is completely shocked during the interaction
with the outer shell, but since not all the kinetic energy is
thermalized in the collision, the shocked regions keep moving after
the end of the collision, producing perturbations in the surface of
the shell. Smaller and/or less dense fragments may be completely
stopped if the shell is sufficiently dense, in which case almost all
the kinetic energy of the motion with respect to the shell is radiated
away.

To further illustrate the above considerations, let us compute the
flare energy and time scale for a shocked fragment with density and
thickness similar to those of the shell: $l_{\mrm{f}} \approx
l_{\mrm{s}} \approx 10^{11}$~cm and $n_{\mrm{f}} \approx n_{\mrm{s}}
\approx 10^{12}$~\percc, where the shell parameters $l_{\mrm{s}}$ and
$n_{\mrm{s}}$ are taken from Terlevich \etal (1995).  The kinetic
energy in the shell frame is $E^{\mrm{f}}_{\mrm{kin}} = 10^{43}
(m_{\mrm{f}} / 10^{-6}) (v_{\mrm{f}} / 10^8)^2$~erg, with
$m_{\mrm{f}}$\ measured in \msun\ and $v_{\mrm{f}}$\ in \cms (here we
switch to characteristic fragment masses comparable to the Earth mass
in order to account for the presence of secondary fragments). Since $a
= \eta = 1$, only half of this energy is actually radiated away during
the flare (eq.~\ref{eq:E_flare}). The radius of a fragment with this
length and density is $R_{\mrm{f}} = 8\ET{13} (m_{\mrm{f}} /
10^{-6})^{1/2}$~cm, so the maximum light crossing time for the flare,
corresponding to a collision at $\theta = 90^\circ$, is about $5200
(m_{\mrm{f}} / 10^{-6})^{1/2}$~s. The crossing and cooling times are
identical for the reverse and forward shocks: $\tau_{\mrm{xing}}
\approx 1500 / (v_{\mrm{f}} / 10^8)$~s and $\tau_{\mrm{cool}} \approx
40 (v_{\mrm{f}} / 10^8)$~s. Earth-mass fragments impacting the shell
at speeds between $1000$\ and $10000$~\kms\ would therefore produce
flares with energies between $5\ET{42}$\ and $5\ET{44}$~erg, and
time-scales between $570$\ and $1540$~s for collisions observed
face-on, or $5770$\ to $6740$~s for collisions observed
edge-on. Fragments ten times more massive would produce flares ten
times more energetic, whose durations would be in the same range as
above for face-on collisions, but larger by a factor of about $3$\ for
edge-on collisions due to the larger radii.  The temperature of the
shocked regions would be in the $0.5$--$50$~keV range for
$v_{\mrm{f}}$\ between $1000$\ and $10000$~\kms. We note that no
strong X~ray absorption is expected, at least for those collisions for
which the cooling time is larger than the shock crossing time, since
in this case the radiation goes through a column of highly ionized gas
(see also Fig.\ 6b in Plewa 1995).

It must be stressed that this simplified scenario for fragment-shell
interaction (thin and perfectly flat shocked fragments colliding
frontally with a thin and perfectly flat outer shell) {\it
overestimates} the efficiency at which the conversion of collision
energy into a high-amplitude X-ray flare proceeds.  Thus, the results
obtained above actually provide a lower limit for the characteristic
time scale and an upper limit for the characteristic amplitude of
X-ray flares in cSNRs, i.e., our estimates are based on the most
favorable conditions to produce short and powerful flares.
\subsection{X-ray light curves and power spectra of 
fragmented cSNRs}
Both our numerical simulations and analytical results indicate that
cSNRs may show substantial X-ray variability originating from the
interactions between the compressed fragments of the ejecta and the
dense outer shell of the remnant. X-ray light curves of cSNRs should
therefore provide a straightforward diagnostic of the presence of
fragments in such objects. In this section we explore the variability
properties associated with fragmented cSNRs by means of Monte Carlo
simulations. These simulations are important not only for comparison
with future X-ray observations of cSNRs, but also to test their
possible connection with active galactic nuclei (AGN). In the
starburst model for AGN (Terlevich \etal 1995 and references therein)
cSNRs are the source of high energy photons and the key ingredient of
the broad line region. While the starburst model proved capable of
reproducing the main spectroscopic and photometric properties of broad
line regions fairly well, it has not so far been able to explain the
rapid X-ray flickering observed in AGN. Our simulations will thus be
useful to determine whether fragment-shell collisions in cSNRs provide
an acceptable model for the rapid X-ray variability of AGN. In fact,
given the lack of X-ray light curves for cSNRs, we shall concentrate
the discussion on this possible application of the fragmented cSNR
model.

The short term cSNR light curve will be the result of a random
superposition of many collision flares, associated with a multitude of
fragments with a variety of masses and sizes interacting with the
shell at different relative velocities and position angles. To compute
light curve simulations we need prescriptions for (1) the time profile
of each flare, (2) the rate of fragment-shell collisions, and (3) the
distribution of fragment properties (sizes, shapes and velocities) at
the moment of collision.

The time-profile of a given flare can be calculated using the
expressions derived above. We take full account of the light travel
time effects, since the light crossing time across the lateral extent
of large fragments can easily exceed the cooling and shock crossing
times for collisions not seen face-on. The most critical assumption in
the computation of the flare time-profile is the hypothesis that the
collision is planar (see below). By assuming a planar collision we are
implicitly limiting our analysis to those {\it sections} of the
fragments which do collide frontally with the shell.

A rough estimate of the number of collisions per unit time ($\nu$) can
be obtained dividing the total number of fragments by the duration of
the shell bombardment phase ($\tau_{\mrm{var}}$), i.e., the phase
during which the cSNR exhibits rapid X-ray variability. If a fraction
$\zeta_{\mrm{f}}$\ of the ejected mass $M_{\mrm{ej}}$\ is in the
form of fragments with an average mass $\ov{m}_{\mrm{f}}$, the mean
rate of collisions is

\beq
\label{eq:collision_rate}
\nu \approx
    \frac{\zeta_{\mrm{f}} M_{\mrm{ej}}}{\tau_{\mrm{var}} \ov{m}_{\mrm{f}}} =
    0.03 \frac{( \zeta_{\mrm{f}} / 0.1)}{(\tau_{\mrm{var}} / {\mrm{yr}})}
         \frac{(M_{\mrm{ej}} / 10 {\mrm{M}_\odot})}{
           (\ov{m}_{\mrm{f}} / 10^{-6} {\mrm{M}_\odot})}
         {\mrm{~s}}^{-1}
\eeq

The distribution of fragment properties at the moment of collision
with the outer shell depends on both the initial distribution and the
details of the evolution of the ensemble of fragments in the shocked
region of the cSNR. Since very little is known about the initial
distribution of sizes and shapes, we are forced to adopt an {\it ad
hoc} parameterization. In the statistical simulations presented below
we assume that masses and densities of the fragments as well as their
radii $R_{\mrm{f}}$ and velocities with respect to the shell have
independent power-law distributions. The shell thickness and density
are kept fixed at $10^{11}$~cm and $10^{12}$~\percc\ respectively.
%
%
\begin{figure*}
\begin{minipage}{17.5cm}
    \protect\centerline{
    \epsfxsize=17.5cm\epsffile[30 400 570 695]{figures/fig5.ps}
     }
\caption{
     {\it Left:} Monte Carlo simulations of the X-ray light curve of
fragmented cSNRs. The numbers in the plots are $\log \nu$\ (in
s$^{-1}$, bottom-left), $\log \ov{L}$\ (in \ergsec, top-right) and the
net r.m.s.\ variability (bottom-right). All light curves are
normalized to the mean luminosity and binned in $500$~s bins. {\it
Right:} Power Spectral Distributions of the light curves.  The power
spectra were computed binning the periodogram using the technique of
Papadakis \& Lawrence (1993). Power-law fits to the
$10^{-5}$--$10^{-3}$~Hz power spectrum are shown; the corresponding
logarithmic slopes are listed in the bottom of the plots.
    \label{fig:LCs}
    }
\end{minipage}
\end{figure*}

Two of the many computed light curves are presented in
Fig.~\ref{fig:LCs}, along with their corresponding power spectral
distribution (PSDs). Both models have fragment masses between
$10^{-8}$\ and $10^{-5}$~\msun, radii between $10^{12}$\ and
$10^{15}$~cm, velocities between $1000$\ and $10000$~\kms\ and
densities between $10^{11}$\ and $10^{13}$~\percc. The {\it
distribution} of these properties is however different in
Figs.~\ref{fig:LCs}a and b: the logarithmic slopes ($\alpha$'s) of
the mass, radius and velocity distributions are $\alpha_{\mrm{m}} =
0$, $\alpha_{\mrm{r}} = 0$\ and $\alpha_{\mrm{v}} = -1$\ in model A
(Fig.~\ref{fig:LCs}a),  and $\alpha_{\mrm{m}} = -1$,
$\alpha_{\mrm{r}} = -2$\  and $\alpha_{\mrm{v}} = 0$\ in model B
(Fig.~\ref{fig:LCs}b), while  fragment densities are homogeneously
distributed ($\alpha_{\mrm{n}} = 0$) in both runs. As a result of
these distribution functions,  model A favors larger and more
massive fragments than model B. Accordingly, the flares in A are
more powerful (with an average energy $\ov{E}_{\mrm{flare}} \approx
3\ET{44}$~erg) and longer ($\ov{\tau}_{\mrm{flare}} \approx
3\ET{4}$~s) than in model B ($\ov{E}_{\mrm{flare}} \approx
5\ET{42}$~erg, $\ov{\tau}_{\mrm{flare}} \approx 7\ET{2}$~s). The
collision rate was  adjusted to 
yield an output mean luminosity of the
order of $10^{42}$~\ergsec. The resulting mass in fragments
impacting the shell per  unit time is $0.6$\ and $3.5$~\msun\peryr\
in  models A and B respectively.  Clearly, collision rates as large
as in model B cannot be sustained for  a long time. (Note that the
smooth background luminosity due to the homogeneous part of the cSNR
is not included in the light curves.  Such a background would have
the effect of increasing the luminosity and diluting the
variations.)

Figs.~\ref{fig:LCs}a and b are examples of runs which produce light
curves whose luminosities, r.m.s.\ variability and power spectra are
similar to those observed in AGN.  Rapidly variable AGN like NGC 4051,
MCG 6-30-15 and NGC 5506 have `red-noise' PSDs with logarithmic slopes
between $-2$\ and $-1$\ in the $10^{-5}$--$10^{-3}$~Hz region
(Lawrence \etal 1987, McHardy \& Czerny 1987, Pounds \& Turner 1987,
McHardy 1989, Green, McHardy \& Lehto 1993, Lawrence \& Papadakis
1993). While some of the simulations have PSDs with logarithmic slopes
in this range (e.g. Figs.~\ref{fig:LCs}a and b), many others do not.

As in any shot-noise model, the PSD is simply the average of the power
spectra of the individual flares (Lehto 1989, Cid Fernandes 1995).
Exponential flares with a decay time $\tau$, for instance, produce a
PSD which is flat at low frequencies bending down to a $f^{-2}$\ law
at high frequencies, the transition taking place within a decade
around $f = 1 / \tau$\ (Lehto 1989). Square pulses with a duration
$\tau$\ produce the same effect, but with additional broad harmonics
in the PSD at multiples of $1 / \tau$. Mathematically, in order to
obtain a logarithmic slope between $-1$\ and $-2$\ over more than two
decades in frequency we need a broad distribution of flare energies
and time-scales such that slow flares dominate the low frequency part
of the PSD while rapid flares dominate the high frequencies. In our
physical model, this spread of energies and time-scales translates
into fragments with different masses, sizes, densities and velocities
colliding with the shell at different angles. Getting the right PSD
slope is a matter of finding a distribution of fragment properties
which results in an appropriate balance of flare energies and
time-scales.  Since many physical quantities are involved in the
definitions of $E_{\mrm{flare}}$\ and $\tau_{\mrm{flare}}$\ there
are many combinations of the $p(m_{\mrm{f}},v_{\mrm{f}},R_{\mrm
{f}},n_{\mrm{f}})$\ distribution function parameters which yield the
same PSD slope, so it is difficult to constrain the parameter space
which yields an AGN-like PSD.  In any case, the simulations show that
at least for some distribution functions the PSDs can be made similar
to that of AGN.

Though the light curve simulations show that the model is in principle
capable of reproducing the observed X-ray variability of AGN, the
physical conditions required to obtain the sharpest ($\tau_{\mrm
{flare}} \sim 1000$~s) and more powerful bursts ($E_{\mrm{flare}} \sim
10^{45}$~erg) seem rather extreme. In order to reach such energies the
fragments must have masses of $10^{-6}$--$10^{-5}$~\msun\ and
velocities of several thousand \kms.  These conditions, together with
the requirement that the cooling and crossing times do not exceed
$1000$~s imply that the fragments should have radii of order
$10^{14}$~cm or more.  Another important constraint is that the shell
is very thin ($l_{\mrm{s}} \sim 10^{11}$~cm), otherwise the shock
crossing times would become exceedingly large.  The strongest
constraint, however, is that the collision with the shell is almost
perfectly planar. Slight distortions of the fragment can easily lead
to flare time-scales much larger than those resulting from a planar
collision. Our hydrodynamic simulations show that the fragments are
far from being smooth, but the resolution we have been able to achieve
is not sufficient for a definitive conclusion.

Highly variable AGNs, as NGC 4051, MCG 6-30-15, and NGC 5506, pose yet
another constraint to the model. In the present scenario it is clear
that variations as large as those observed in these objects cannot be
sustained throughout the cSNR lifetime, otherwise the mass and energy
in the form of fragments would exceed those of the homogeneous part of
the remnant. The simplest solution to this problem would be to
postulate that these objects are in an evolutionary phase where rapid
X-ray variability is at its peak: it either has been weaker in the
past or will be so in the future. This is however in conflict with the
results of Green, McHardy \& Lehto (1993), who found no evidence for
changes in the variability properties over intervals smaller than $3$\
years. It is nevertheless interesting to note that objects like NGC
4151 and NGC 5548 do {\it not} show variations as large as NGC 4051,
MCG 6-30-15 or NGC 5506, despite their similar X-ray luminosities
(Green, McHardy \& Lehto).

To conclude, fragment-shell collisions are capable of generating
`chaotic' X-ray variability in cSNRs. With the free parameters
suitably adjusted, X-ray light curves similar to those observed in AGN
can be obtained. Our feeling, however, is that in the case of extreme
variability as that observed in AGN like NGC 4051, MCG 6-30-15 and NGC
5506 the present model is strained too much.  On the other hand, it
seems definitely worthwhile to launch a program of X-ray monitoring of
cSNRs, not only to test the hydrodynamic effects predicted here, but
to provide a strong empirical test of the possible connection between
cSNRs and AGN.
\section{Conclusions}
We have studied the evolution of inhomogeneities (fragments) in
supernova ejecta as the remnant interacts with a dense medium. Under
such conditions, the ejected fragments, particularly the denser ones,
experience a complex evolution and end up colliding with the remnant
outer shell.  The calculations reported in this paper show that:
  \begin{itemize} 
  \item{} Fragments are prone to further fragmentation inside the
shocked region of the remnant. It is conceivable, however that at
least some fragments (those larger and denser) in advanced
evolutionary phases of the remnant (when the distance between the
forward and reverse shocks begins to decrease) may approach the outer
shell as well defined entities.
  \item{} The shocked fragments eventually collide with the outer shell at
velocities reaching several thousands \kms. As both the fragments and
the shell are already strongly compressed due to intense cooling,
their densities at the moment of collision are very high (up to some
$10^{12}$ \percc), and the shocks generated by the collision
are strongly radiative. The outcome of such an energetic collision is
an X-ray flare.
  \item{} The observed total energy of the flare is essentially
determined by the kinetic energy of the fragment in the reference
frame of the shell. The observed duration of the flare depends both on
shell and fragment geometries and on the inclination of the collision
plane with respect to the observer's line of sight. Under the most
favorable conditions (a head-on collision of an entirely flat fragment
with an entirely flat shell, collision plane perpendicular to the line
of sight) the collision energy can be thermalized and radiated in a
fraction of an hour.
  \item{} Fragment-shell collisions occurring randomly in space and
time are in principle capable of generating light curves and power
spectra similar to those observed in AGNs, but the conditions required
to reproduce the sharpest and most powerful bursts seem contrived.
  \end{itemize}

On the theoretical side, future research should explore modifications
of the basic scenario outlined in the present paper.  For example, it
is conceivable that X-ray flares can also be produced by collisions
between ejected fragments and circumstellar clouds (see McKee
1983). The external cloudlets entering the hot shocked region would
evolve similarly to the ejecta fragments, and a cloud-fragment
collision should be similar to a fragment-shell collision. Another
area for exploration is the interaction between ejecta fragments and
the reverse shock thin shell. In the vicinity of the reverse shock,
given that even low-$\epsilon$ fragments are not yet severely
distorted, the probability of a cloud--fragment collision is
increased. Also, because of the resulting larger relative velocities,
one would expect stronger radiation flashes and with a shorter
duration than those from collisions of ejecta fragments with the main
shell.

Further simulations with higher resolution will be necessary to
uncover the details of fragment-shell collision. Length scales of the
order of $10^{12}$ and $10^{14}$~cm are needed to produce sharp X-ray
flares. The corrugation of the shell in Fig. \ref{fig:caseD} has a
characteristic wavelength of $\sim3\times10^{15}$~cm), which is the
expected value for the fastest growing mode of the thin shell
instability in our simulation (the fastest growth rate appears on
scales comparable to the thickness of the shell; Vishniac 1993). Thus,
one can expect that an increase in numerical resolution would lead to
a shell which is thinner and denser (see Sect 2.4), and with a finer
corrugation scale. With our present computational resources, the
required resolution cannot be achieved but, given the rapid
improvement in speed and memory resources, the problem could be
approached in a few years.

On the observational side, the basic verifiable prediction of the
fragmented ejecta model is that cSNRs by themselves should exhibit
stochastic and rapid X-ray variability. 
Unfortunately, this prediction
is also hard to test because {\it bona fide} cSNRs are hard to come by,
being generated at unpredictable times and places. To date, the
only relevant observation is that of the cSNR at the core of M~82
(Collura \etal 1994), which indeed show variations in scales of a
day. Ideally, one would like to monitor a few cSNRs for a few years,
but this would require large amounts of satellite time. SN~1988Z
(Stathakis \& Sadler, Turatto \etal 1993) recently observed as a
strong X-ray source by ROSAT (Fabian \& Terlevich 1996) is an obvious
first target for such a project.
\section*{Acknowledgments}
We would like to thank Andy Fabian, Martin Rees, the referee - Robin
Williams for important suggestions. RCF acknowledges the Brazilian
institution CAPES for grant 417/90-5. The work of MR was supported by
the grant KBN 2P-304-017-07 from the Polish Committee for Scientific
Research. MR also thanks the Instituto de Astronom\'{\i}a--UNAM for
its hospitality. JF thanks the hospitality of the Copernicus
Astronomical Center, and acknowledges the support given to this
project by DGAPA-UNAM grant IN105894, CONACyT grant 400354-5-4843E,
and by a R\&D CRAY Research grant. JF and GTT acknowledge partial
support from the EEC grant for international collaboration
CI1*-CT91-0935.  The work of WM was supported by the National
Aeronautic and Space Administration under grants number NAG5-629 and
NAGW-2532. The simulations were performed on workstation cluster at
the Max-Planck-Institut f\"ur Astrophysik and on the Cray Y-MP at the
Supercomputing Center--UNAM (Mexico).
\bsp
\label{lastpage}

\begin{thebibliography}{99}
%
%
%
\bibitem[\protect\citename{ }19]{}
	Anderson M. C., Jones T. W., Rudnick L., Tregilis I. L., Kang H., 1994,
	ApJ, 421, L31
\bibitem[\protect\citename{ }19]{}
	Arnett D., 1988,
	ApJ, 331, 377
\bibitem[\protect\citename{ }19]{}
	Arnett D., Fryxell B., M\"uller E., 1989,
	ApJ, 341, L63
\bibitem[\protect\citename{ }19]{}
	Aschenbach B., Egger R., Tr\"umper J., 1995,
	Nature, 373, 587 
\bibitem[\protect\citename{ }19]{}
	Bandiera R., van den Bergh S., 1991,
	ApJ, 374, 186
\bibitem[\protect\citename{ }19]{}
	Berger M. J., Colella P., 1989,
	J. Comput. Phys., 82, 64
\bibitem[\protect\citename{ }19]{}
	Berger M. J., Oliger J., 1984,
	J. Comput. Phys., 53, 484
\bibitem[\protect\citename{ }19]{}
	Boller T., Fink H., Schaeidt S., 1994,
	A\&A, 291, 4
\bibitem[\protect\citename{ }19]{}
	Braun R., Gull S. F., Perley R., 1987,
	Nature, 327, 395
\bibitem[\protect\citename{ }19]{}
	Chevalier R. A., 1982,
	ApJ, 258, 790
\bibitem[\protect\citename{ }19]{}
	Chugai N. N., 1993,
	ApJ, 414, L101
\bibitem[\protect\citename{ }19]{}
	Chugai N. N., Danziger I. J., 1994,
	MNRAS, 268, 173
\bibitem[\protect\citename{ }19]{}
	Cid Fernandes R., Terlevich R., 1994,
	in Tenorio-Tagle, G., ed.,
	Violent Star Formation: from 30 Doradus to QSOs.
	Cambridge University Press, Cambridge, p. 365
\bibitem[\protect\citename{ }19]{}
	Cid Fernandes R., 1995,
	PhD Thesis (Part 2), Institute of Astronomy, University of Cambridge, UK.
\bibitem[\protect\citename{ }19]{}
	Colella P., Woodward P.R., 1984,
	J. Comput. Phys., 59, 264
\bibitem[\protect\citename{ }19]{}
	Collura A., Reale F., Schulman E., Bregman J., 1994,
	ApJ, 420, L63
\bibitem[\protect\citename{ }19]{}
	Draine B.T., Woods D. T., 1991,
	ApJ, 383, 621
\bibitem[\protect\citename{ }19]{}
	Fabian, A. C., Terlevich, R., 1996,
	MNRAS (submitted)
\bibitem[\protect\citename{ }19]{}
	Fesen R. A., Matonick D. A., 1993,
	ApJ, 407, 110
\bibitem[\protect\citename{ }19]{}
	Franco J., Ferrara A., R\'o\.zyczka M., Tenorio-Tagle G., Cox D. P., 1993a,
	ApJ, 407, 100
\bibitem[\protect\citename{ }19]{}
	Franco J., Melnick J., Terlevich R., Tenorio-Tagle G., R\'o\.zyczka M., 1993b,
	in Franco J., Ferrini F., Tenorio-Tagle G., eds,
	Star Formation, Galaxies and the Interstellar Medium.
	Cambridge University Press, Cambridge, p. 149
\bibitem[\protect\citename{ }19]{}
        Franco, J., Miller, W., Cox, D., Terlevich, R., Tenorio-Tagle, G.
1993c, 
        Rev. Mex. Astron. Astrofis., 27, 133 
\bibitem[\protect\citename{ }19]{}
        Franco J., Miller W., Arthur S. J., Tenorio-Tagle G., Terlevich R.
1994, 
        ApJ, 435, 805
\bibitem[\protect\citename{ }19]{}
	Franco J., Tenorio-Tagle G., Bodenheimer P., R\'o\.zyczka M., 1991,
	PASP, 103, 803
\bibitem[\protect\citename{ }19]{}
	Fryxell B., 1994,
	in Franco J., Lizano S., Aguilar L., Daltabuit E.,
	Numerical Simulations in Astrophysics.
	Cambridge University Press, Cambridge, p. 175
\bibitem[\protect\citename{ }19]{}
	Green A. R., McHardy I. M., Lehto H. J., 1993,
	MNRAS, 265, 664
\bibitem[\protect\citename{ }19]{}
	Hachisu I., Matsuda T., Nomoto K., Shigeyama T., 1992,
	ApJ, 390, 230
\bibitem[\protect\citename{ }19]{}
	Hamilton A. J. S., 1985,
	ApJ, 291, 523
\bibitem[\protect\citename{ }19]{}
	Hanuschik R. W., Spyromilio J., Stathakis R., Kimeswenger S., Lasker B. M., Golinowski D. A., 1991,
	ApJ, 371, 563
\bibitem[\protect\citename{ }19]{}
	Kimura, T., Tosa, M., 1991,
	MNRAS, 251, 664
\bibitem[\protect\citename{ }19]{}
	Lawrence A., Pounds K. A., Watson M. G., Elvis M. S., 1987,
	Nature, 325, 692
\bibitem[\protect\citename{ }19]{}
	Lawrence A., Papadakis I., 1993,
	ApJ, 414, L85
\bibitem[\protect\citename{ }19]{}
	Lehto H. J., 1989,
	in Hunt J., Bottrick B., eds,
	Proceedings of the 23rd ESLAB Symposium.
	ESA publication Division, p. 499
\bibitem[\protect\citename{ }19]{}
	Lucy L. B., Danziger I. J., Gouiffes C., Bouchet P., 1989,
	in Tenorio-Tagle G., Moles M., Melnick J., eds,
	Structure and Dynamics of the ISM.
	Springer, Berlin, p. 164
\bibitem[\protect\citename{ }19]{}
	Klein R. I., McKee C. F., Colella P., 1994,
	ApJ, 420, 213
\bibitem[\protect\citename{ }19]{}
	Kronberg P. P., Biermann P., Schwab F. R. 1985, 
	ApJ, 291, 693
\bibitem[\protect\citename{ }19]{}
	McHardy I. M., 1989,
	in Hunt J., Bottrick B., eds,
	Proceedings of the 23rd ESLAB Symposium.
	ESA publication Division, p. 1111
\bibitem[\protect\citename{ }19]{}
	McHardy I., Czerny B., 1987,
	Nature, 325, 696
\bibitem[\protect\citename{ }19]{}
	McKee C. F., 1983,
	in Danziger J., Gorenstein P., eds,
	Supernova Remnants and Their X-ray Emission (IAU Symp. 101).
	Reidel, Dordrecht, p. 87
\bibitem[\protect\citename{ }19]{}
	M\"uller E., Fryxell B., Arnett D., 1991,
	A\&A, 251, 505
\bibitem[\protect\citename{ }19]{}
	Papadakis I. E., Lawrence A., 1993a,
	MNRAS, 261, 612
\bibitem[\protect\citename{ }19]{}
	Papadakis I. E., Lawrence A., 1993b,
	Nature, 361, 233
\bibitem[\protect\citename{ }19]{}
	Plewa T., 1995,
	MNRAS, 275, 143
\bibitem[\protect\citename{ }19]{}
	Plewa T., M\"uller E., 1996,
	Comp. Phys. Commun. (submitted)
\bibitem[\protect\citename{ }19]{}
	Pounds K. A., Turner T. J., 1987,
	in Treves A., ed.,
	Variability of Galactic and Extragalactic X-ray Sources.
	Associazione per l'avanzamento dell'astronomia, Bologna, p. 1
\bibitem[\protect\citename{ }19]{}
	Seward F., Gorenstein P., Tucker W., 1983,
	ApJ, 266, 287
\bibitem[\protect\citename{ }19]{}
	Shull M., 1980,
	ApJ, 237, 769
\bibitem[\protect\citename{ }19]{}
	Stathakis R. A., Sadler E. M., 1991,
	MNRAS, 250, 786
\bibitem[\protect\citename{ }19]{}
	Strom R., Johnston H. M., Verbunt F., Aschenbach B., 1995,
	Nature 373, 590 
\bibitem[\protect\citename{ }19]{}
	Tenorio-Tagle, G., 1994,
	in Clegg R. E. S., Stevens I. R., Meikle W. P. S., eds, 
	Circumstellar Media in the Late Stages of Stellar Evolution.
	Cambridge University Press, Cambridge, p. 166
\bibitem[\protect\citename{ }19]{}
	Tenorio-Tagle G., R\'o\.zyczka M., Franco J., Bodenheimer P., 1991,
	MNRAS, 251, 318
\bibitem[\protect\citename{ }19]{}
	Terlevich R., 1994,
	in Clegg R. E. S., Stevens I. R., Meikle W. P. S., eds, 
	Circumstellar Media in the Late Stages of Stellar Evolution.
	Cambridge University Press, Cambridge, p. 153
\bibitem[\protect\citename{ }19]{}
	Terlevich R., Tenorio-Tagle G., Franco J., Melnick J., 1992,
	MNRAS, 255, 713
\bibitem[\protect\citename{ }19]{}
	Terlevich R., Tenorio-Tagle G., R\'o\.zyczka M., Franco J., Melnick J., 1995,
	MNRAS, 272, 198
\bibitem[\protect\citename{ }19]{}
	Turatto M., Capellaro E., Danziger I. J., Benetti S., Gouiffes C., Della Valle M., 1993,
	MNRAS, 262, 128
\bibitem[\protect\citename{ }19]{}
	Vishniac E.T., 1983,
	ApJ 274, 152
\bibitem[\protect\citename{ }19]{}
	Wang L., Hu J., 1994,
	Nature, 369, 380
\bibitem[\protect\citename{ }19]{}
	Wheeler J. C., Mazurek T. J., Sivaramakrishnan A., 1980,
	ApJ, 237, 781
\bibitem[\protect\citename{ }19]{}
	Winkler P. F., Tuttle J. H., Kirshner R. P., Irwin M. J., 1988,
	in Roger R. S., Landecker T. L., eds,
	Supernova Remnants and the Interstellar Medium (IAU Coll. 101).
	Cambridge University Press, Cambridge, p. 65
%
%
%
\end{thebibliography}
\end{document}